\documentclass{ws-procs975x65}

\begin{document}

\title{The Deceptively Boring PSR J1738+0333}

\author{PAULO FREIRE$^*$ \& NORBERT WEX}

\address{Max-Planck-Institut f\"ur Radioastronomie\\
Auf dem H\"ugel 69, Bonn, 53121, Germany\\
$^*$E-mail: pfreire@mpifr-bonn.mpg.de}

\begin{abstract}
We present preliminary results of 7 years of Arecibo timing of the
pulsar-white dwarf binary PSR J1738+0333. We can measure the proper motion,
parallax with excellent precision and have detected the orbital decay.
Furthermore, the companion has been detected at optical
wavelengths and a mass ratio of $8.1 \pm 0.3$ has been measured from the
orbital variation of its Doppler shift. Once the companion mass is determined
from the optical measurements, this system will provide strong limits for
the radiation of dipolar gravitational waves. Assuming that general relativity
holds, the fast-improving measurement of the orbital decay, combined
with the measurement of the mass ratio, will provide an independent
and precise measurement of the component masses.
\end{abstract}


\bodymatter

\section{Introduction}

PSR~J1738+0333 is a 5.85\,ms pulsar in a low eccentricity binary system
with orbital period of $P_b\,=\, 8.5\,\rm h$ \cite{jbo+07} and a low mass
helium white dwarf (WD) companion. The pulsar was discovered in 2001 in
a 20-cm intermediate Galactic latitude survey \cite{jbo+07} carried out with
the multi-beam system of the Parkes Telescope. Given its low Northern
declination, it can be observed with much improved sensitivity
using the the 305-m Arecibo Telescope. For that reason it has been
timed with Arecibo since 2003 using the L-wide receiver (with frequency
coverage from 1100 to 1700 MHz). We used the Wide-band Arecibo Pulsar Processor
(WAPP) correlators and the Arecibo Signal Processor
(a coherent dedispersion machine) as back-ends. The WAPPs have
bandwidths of 100 MHz centered at 1170, 1410 and 1510 MHz (and, occasionally,
a fourth WAPP centered at 1310 MHz as well).

The WAPP correlator data were Fourier transformed to obtain one 256-point
power spectrum every 64 $\mu$s. These are then folded and dedispersed
at the pulsar's spin period and DM using
``sigfoldpow'', a program designed by Ingrid Stairs specifically for
producing highly accurate average pulse profiles from search data.
The ASP data have not yet been fully analyzed, that is one of the reasons
why these results are still preliminary.
The observed pulse profiles are then cross-correlated with a low-noise
template, the best fit yields a pulse time of arrival at the telescope
(TOA). The TOAs were then analysed using the 
TEMPO\footnote{\url{http://www.atnf.csiro.au/research/pulsar/tempo/}} timing
software, which makes a least-squared fit of the observed barycentric
TOAs to a timing model, which includes spin, astrometric and orbital
parameters.

\section{Results}

We have obtained a TOA residual rms of 1.4 $\mu$s, with TOAs taken every
268 s for each of the 12 (or 16) simultaneous 12.5-MHz sub-bands.
This results in more than 16000 TOAs. Of these,
we use 9000 with a stated accuracy better than 3 $\mu$s. This
large number of precise TOAs provide extraordinarily precise timing
parameters. These include the proper motion
($\mu_{\alpha} = 7.041 \pm 0.007\, \rm mas \, yr^{-1}$ and
$\mu_{\delta} = 5.135 \pm 0.019 \, \rm mas \, yr^{-1}$) and parallax
($\pi = 0.75 \pm 0.05 \, \rm mas$), which corresponds to
a distance of $d = 1.33 \pm 0.09$\,kpc.
The orbital period is measured so precisely
($P_b = 30653.9194513 \pm 0.0000005$ s) that we can already
detect the orbital decay, albeit to only 2-$\sigma$ precision:
$\dot{P_{b}}^{\rm obs}= (-2.1 \pm 1.0) \times 10^{-14} \rm \, s\,s^{-1}$.
This becomes more significant after subtracting the extrinsic
orbital period variation caused by
the proper motion of the system \cite{shk70} and the
Galactic acceleration:
$\dot{P_{b}}^{\rm ext}= (0.73 \pm 0.08) \times 10^{-14} \rm \, s\,s^{-1}$.
The intrinsic orbital period derivative is
$\dot{P_{b}}^{\rm int}= (-2.9 \pm 1.0) \times 10^{-14} \rm \, s\,s^{-1}$.

\section{Optical Observations}


The companion white dwarf has been detected, and it has deep
spectral lines typical of a $\sim 0.2\,M_{\odot}$ thick-envelope
He white dwarf. These were used to determine a radial-velocity curve
using Gemini South on Cerro Pach\'on.
From this we can derive the mass ratio of the system, $R = 8.1 \pm 0.3$
(Van Kerkwijk, 2007, pers. comm.). At the moment we don't have
a good estimate of the companion mass. However, if it follows the
relation postulated by Tauris \& Savonije in 1999\cite{ts99},
then the companion mass $m_c \simeq\,0.18\,\pm\,0.02\,M_{\odot}$, which
is qualitatively consistent with the observed spectral properties.
Given the mass ratio this results in a pulsar mass
$m_p \simeq\,1.46\,\pm\,0.16\, M_{\odot}$
and an orbital inclination $i \simeq 33^\circ$. Apart from resulting in a
reasonable value for $m_p$, the low inclination is consistent with the
non-detection of the Shapiro delay. Furthermore, when we introduce
Shapiro delay parameters $m_c$ and $\sin i$ in our solution, the apparent
eccentricity diminishes to an insignificant level of $0.00000039 \pm
0.00000013$.
This implies that the difference between the semi-major and semi-minor
axes is only $8 \pm 6 \, \mu \rm m$! 

\section{Limit on dipolar gravitational wave emission}

\begin{figure}[t]
\begin{center}
\psfig{file=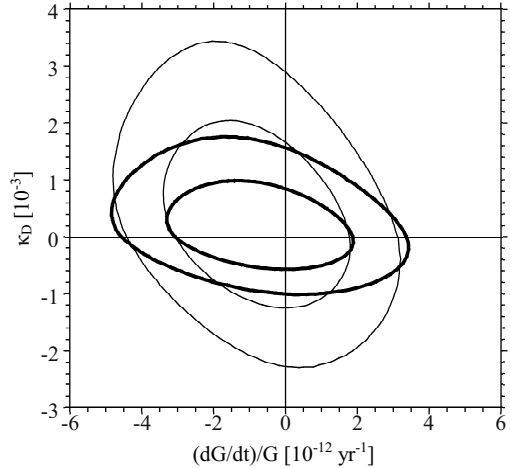,width=2.8in}
\end{center}
\caption{Limits on $\dot{G}$ and $\kappa_{D}$ derived from the joint analysis
of the timing data for PSR~J0437$-$4715 plus PSR~J1012+5307 (thin contours)
or PSR~J1738+0333 (thick contours). The inner contours include 68.3\%
and the outer contours include 95.4\% of the simulated probability in each case.
}
\label{fig:limits}
\end{figure}

If these values for $m_c$ and $m_p$ hold, the
orbital decay due to the emission of quadrupolar
gravitational waves predicted by general relativity is 
$\dot{P_{b}}^{\rm GR} = -2.7 \times 10^{-14} \rm \, s\,s^{-1}$.
This is consistent with the intrinsic orbital decay
$\dot{P_{b}}^{\rm int}$.

Subtracting $\dot{P_{b}}^{\rm GR}$ from $\dot{P_{b}}^{\rm int}$
we obtain an ``excess'' of
$\dot{P_{b}}^{\rm exc} = (-0.2 \pm 1.0 ) \times 10^{-14} \rm \, s\,s^{-1}$.
Following the logic of Lazaridis et al. (these proceedings),
we can combine this ``excess'' with that from a pulsar with a longer
orbital period, like PSR~J0437$-$4715
\cite{vbv+08}, and derive joint limits for the variation of
the gravitational constant $\dot{G}$ and the dipolar gravitational
wave emission constant $\kappa_{D}$.
In figure~\ref{fig:limits}, we can see that, provided that
the companion mass estimate is realistic, PSR~J1738+0333 already
provides excellent limits on dipolar gravitational wave
emission:
$\kappa_{D} = 1.4^{+5.6}_{-4.7} \times 10^{-4}$ (1-$\sigma$) and
$\kappa_{D} = 1.4^{+13.3}_{-9.3} \times 10^{-4}$ (2-$\sigma$). This
happens despite it having a baseline that is only
half that of PSR~J1012+5307: this is more than compensated
by its shorter orbital period and better timing
precision. This also means that the measurement of $\dot{P_{b}}^{\rm int}$
will keep improving much faster for PSR~J1738+0333 in the
future. One of the consequences is that, if we assume
general relativity to hold (i.e., $\dot{P_{b}}^{\rm int} =\dot{P_{b}}^{\rm GR}$),
we will obtain precise values for
$m_c$, $m_p$ and $i$ from the measurements of $\dot{P_{b}}^{\rm GR}$
and $R$. These will help to calibrate the spectral models of
low-mass WDs.



\begin{thebibliography}{10}
\bibitem{jbo+07} Jacoby, B.~A., Bailes, 
M., Ord, S.~M., Knight, H.~S., \& Hotan, A.~W.\ 2007,
{\em Ap.J.}, {\bf 656}, 408 

\bibitem{shk70}
I.~S. Shklovskii, {\em Soviet Ast.} {\bf 13}, 562 (1970).

\bibitem{ts99} Tauris, T.~M., \& Savonije, G.~J.\ 1999, {\em A\&A}, {\bf 350}, 928 

\bibitem{lwj+09}
K.~Lazaridis, N.~Wex, A.~Jessner and et~al., {\em MNRAS} {\bf 400}, 805 (2009).

\bibitem{vbv+08}
J.~P.~W. Verbiest, M.~Bailes, W.~van Straten and et~al., {\em ApJ} {\bf 679},
  675 (2008).
\end{thebibliography}

\end{document}